\documentclass[aps,prd,twocolumn,groupedaddress,showpacs]{revtex4-1}

\usepackage{verbatim} 
\usepackage{xcolor}
\usepackage[utf8]{inputenc}
\usepackage{bbold}
\usepackage{graphicx}
\usepackage{float}
\usepackage{amsmath}
\usepackage{mathrsfs}

\graphicspath{{fig/}}
\usepackage{epstopdf}

\usepackage{bbm}

\usepackage[toc]{appendix}
\usepackage[normalem]{ulem}

\begin{document}

\title{Fluctuations of differential number counts of radio continuum sources}

\author{Song Chen}
\email{songchen@physik.uni-bielefeld.de}

\author{Dominik J. Schwarz}
\email{dschwarz@physik.uni-bielefeld.de}

\affiliation{Fakult\"{a}t f\"{u}r Physik, Universit\"{a}t Bielefeld, Postfach
100131, 33501 Bielefeld, Germany}


\date{\today}

\begin{abstract}
We investigate the differential number counts of sources in radio continuum surveys, including 
all terms at linear order in cosmological perturbations. Our framework does not assume a specific 
gauge condition. This general approach allows us to recover gauge invariance explicitly. 
With the complete derivations of the covariant volume integral on
the past light cone, we have identified several contributions in the number
counts. To clarify their underlying physics, we present each contribution in terms of scalar, vector 
and tensor modes. This theoretical framework promises to be widely applicable to continuum 
radio galaxy surveys to model the expected angular power spectrum and two-point correlation.
\end{abstract}


\maketitle

\section{\label{sec:1} Introduction}

Number counts of extragalactic radio continuum sources were one of the first
cosmological probes, and allowed to test the evolution of the Universe 
well before the discovery of the cosmic microwave background (CMB).

Historically, they have been crucial to dismiss the steady state model of cosmology, 
falsify the so-called perfect cosmological principle, and to establish the isotropy of the Universe 
at large angular scales (see e.g.~\cite{Ryle1961,RyleClarke, Longair1966}). 
In those early days, the number of radio sources was only of the order of a few thousand, which 
allowed a rough test of the isotropy of the Universe, but the investigation of the small 
count fluctuations expected at large angular scales was dominated by shot noise and systematics.   

Upcoming radio continuum surveys from a new generation of radio interferometers, such as the 
Low Frequency Array  (LOFAR) \footnote{URL: www.lofar.org}, the Australian Square Kilometre 
Array Pathfinder (ASKAP) \footnote{URL: www.atnf.csiro.au/projects/askap/} and the 
Square Kilometre Array (SKA) \footnote{URL: www.skatelescope.org} will touch unprecedented large 
survey volumes and flux ranges. Therefore the catalogs emerging from these surveys will no longer be 
limited by small numbers, on the contrary they will compete with and outreach the biggest extragalactic 
source counts in other wave bands. Consequently, accurate theoretical modeling of radio source number 
counts will be crucial to understanding the underlying physics.

Extragalactic radio sources are diverse in nature and evolve both in comoving number density and 
luminosity function (see e.g.~\cite{Zotti2010}). They fall into two classes of objects: active galactic nuclei 
(AGN) and star forming galaxies (SFG). The angular resolution of SKA continuum surveys will allow to 
classify the sources according to their morphology. 

Active galactic nuclei are the brightest sources in radio continuum surveys.  Their radio emission
is due to synchrotron radiation emerging from the vicinity of their central supermassive objects, presumably 
black holes. AGN is common from the local Universe out to redshifts of $z\sim7$, 
are distributed over the whole sky and are extremely luminous, especially at low frequencies. 
 
This encourages us to investigate them for large scale structure and cosmology. 
This direction has been explored previously by several authors ~\cite{Blake2002,Blake2004a}
based on data from the NRAO VLA Sky Survey  ~\cite{Condon1998}. 
However, in the interpretation of the data only the density perturbations itself, but no effects 
of light propagation have been considered.

In this work we provide the theoretical basis to calculate the differential number counts
\begin{equation}
{{\rm d}^2 N \over {\rm d} \Omega {\rm d} S} ({\bf \hat e}, \omega, S), 
\end{equation} 
which denotes source number per solid angle and per flux density 
observed in direction ${\bf \hat e}$ in a narrow frequency 
band centered at frequency $\omega$ and at flux density $S$.
In contrast to galaxy redshift surveys, the distance estimates of the sources have to rely on the
observed brightness and thus on the luminosity distances. In radio, the synchrotron and free-free 
emission mechanisms suggest that the specific luminosity of radio sources should follow a power law 
$L(\omega)\propto \omega^{-\alpha}$, where $\alpha$ is named the spectral index. Our results are not limited to the 
radio band, with proper K-correction they hold for any flux-limited sample obtained in a narrow 
frequency band.

The linear order effects in the number of galaxies per redshift per solid angle was investigated in 
~\cite{Yoo2009a,Challinor2011,Yoo2009} for different choices of coordinates (gauges).
A more general approach was presented in ~\cite{Bonvin2011,Yoo2010} without specifying any gauge 
condition. These results are most significant for optical galaxy redshift surveys like 
BOSS \footnote{URL: www.sdss3.org/.} (LRG $z < 0.7$) and Euclid \footnote{URL: sci.esa.int/euclid.} ($z < 2$).
Compared to the analysis of optical galaxies, the investigation of radio continuum sources should 
put more focus on the distortion effects at higher redshifts. So far no fully relativistic treatment 
for the differential number counts of radio sources is available. 

In this work, we provide the complete theoretical framework of differential number counts of radio 
sources at linear order in cosmological perturbation theory. 
In our results, part of the perturbations to the differential number counts have been investigated in ~\cite{Maartens:2012rh,Raccanelli}, where they choose the Newtonian gauge.
Because of the inherent gauge freedom in the general relativity perturbation theory, the gauge choice of the 
perturbations is always an issue especially on super-Hubble scales.
We do not make any gauge assumption to ensure that their physical meanings can be extracted clearly.

The paper is structured as follows. In the next section, we show how to count objects on the past light cone.
In Sec. \ref{sec:OB} we express this counting in the observed coordinates, where we derived the flux fluctuation to the linear order.
The total volume distortion including the flux distortion and angular displacements can be seen clearly.
Finally, we combine the luminosity function and previous number counts elements into the first order number count per flux per solid angle.

\section{\label{sec:2} Number Counts on the light cone}

In this work, we consider linear perturbations of a spatially isotropic, homogeneous and flat metric, largely
following the notation of ~\cite{Mukhanov}. A dot denotes a derivative with respect to the conformal time 
$\eta$, the scale factor is $a$ and ${\cal H} \equiv \dot{a}/a$. The line element is expressed as
\begin{eqnarray}\label{eq:metric}
 {\rm d} s^2&=&-a^2(1+2\phi){\rm d}\eta^2-2a^2(B_{,i}+S_i ){\rm d}\eta {\rm d}x^i \\
 &&+a^2[(1+ 2\psi) \delta_{ij} +2E_{,ij} +F_{i,j}+F_{j,i}+h_{ij}]{\rm d}x^i{\rm d}x^j,\nonumber
\end{eqnarray}
where $B_{,i}= \partial B/\partial x^i$, and $S_i$ and $F_i$ are transverse vectors, 
i.e.~their divergencies vanish ($S^i_{,i}=0$ and $F^i_{,i}=0$).
The transverse, traceless tensor $h_{ij}$ satisfies the four constraints $h^i_i=0, h^i_{j,i}=0$. We express our 
results in terms of the gauge invariant metric potentials
\begin{eqnarray}
&&\Phi\equiv\phi- \mathcal{H}(B+\dot{E})-(\dot{B}+\ddot{E}), \nonumber \\ 
&&\Psi\equiv\psi-\mathcal{H}(B+\dot{E}), \\
&&U_i\equiv 	S_i+\dot{F}_i. \nonumber
\end{eqnarray}
We can consider our past light cone to be a three-dimensional hypersurface of
the four-dimensional space-time ~\cite{S.Weinberg}. 
Within this hypersurface, the four coordinates $x^\mu$ may be expressed by 
smooth functions of three parameters $y^\alpha$: 
\begin{equation}
 x^\mu=x^\mu(y^1,y^2,y^3).
\end{equation}
For convenience, we use the light cone constraint to fix the conformal time
$\eta$, and choose the three parameters on the past light cone to be
the spherical coordinates $(r,\theta,\phi)$. In a second step (next section) we connect them to 
the observed source positions on the sky and to observed comoving source distances.

The total number of radio sources on the past light cone (plc) can be computed by
considering a covariant volume integral
\begin{equation}
N = \int_{\rm plc}  n_{\rm phy} u^\mu {\rm d} S_\mu, 
\end{equation}
where $n_{\rm phy} = n_{\rm phy}(\eta,x^i)$ is the inhomogeneous physical number density in the rest frame of the cosmic fluid, 
$u^0 = (1-\phi)/a, u^i = v^i/a$ are the components of the four-velocity field of the 
radio sources and
\begin{eqnarray}
 {\rm d}S_\mu&=&\epsilon_{\mu\nu\sigma\rho} {\rm d}x^\nu {\rm d}x^\sigma {\rm d} x^\rho \nonumber \\
 &=&\epsilon_{\mu\nu\sigma\rho}\frac{\partial x^\nu}{\partial r}\frac{\partial
x^\sigma}{\partial \theta}\frac{\partial x^\rho}{\partial \varphi} {\rm d} r {\rm d} \theta
{\rm d} \varphi,
 \end{eqnarray}
with $\epsilon_{\mu\nu\sigma\rho}=\sqrt{-g}[\mu\,\nu\,\sigma\,\rho]$ 
denoting the Levi-Civita pseudotensor.

At linear order, the covariant volume integral can be written as
\begin{eqnarray}\label{eq:covolume}
 N&=&\int_{\rm plc}  n_{\rm phy} u^\mu  
\epsilon_{\mu\nu\sigma\rho}\frac{\partial x^\nu}{\partial r}\frac{\partial
x^\sigma}{\partial \theta}\frac{\partial x^\rho}{\partial \varphi}  {\rm d} r {\rm d} \theta
{\rm d} \varphi \nonumber \\
&=&\int_{\rm plc} n_{\rm phy}  a^3 r^2 [1+3\psi+\nabla^2 E+v^i e^r_i ]{\rm d} r {\rm d} \Omega,
\end{eqnarray}
where $e^r_i$ denotes the radial unit vector.  
Here, the terms $3\psi$ and $\nabla^2 E$ are due to the distortion of the spatial volume,
the term $v^i e^r_i$ is due to the light cone projection.
Let us stress that this result holds true for all coordinate systems in which the observer is at rest, i.e. $v^i$ denote the velocity of the sources. 
In order to express the $N$ in another frame (e.g. the CMB rest frame) one has to replace $v^i$ by $(v^i-v_o^i)$ where $v_o^i$ denotes the observer's peculiar velocity. 
This can be easily seen from the fact that linearized Lorentz boost reduce to Galilean transformations which do not modify the volume, but affect the light cone projection.
By construction $N$ is a gauge invariant quantity, which we have checked explicitly.

\section{\label{sec:OB}Coordinates of the observer}

In the previous section, the number count has been expressed as an integral over the
coordinates $(r,\theta,\varphi)$. However, these coordinates do not agree with the coordinates 
used by the observer. The actual observables are redshift and/or flux, instead of coordinate distance,
and position (two observed angles), instead of the angular coordinates introduced above.
The following subsection briefly reviews the redshift and luminosity distance distortions up to first order in 
cosmological perturbations.

\subsection{Redshift distortions}

\begin{figure}
\centering
 \includegraphics[width=0.9\linewidth]{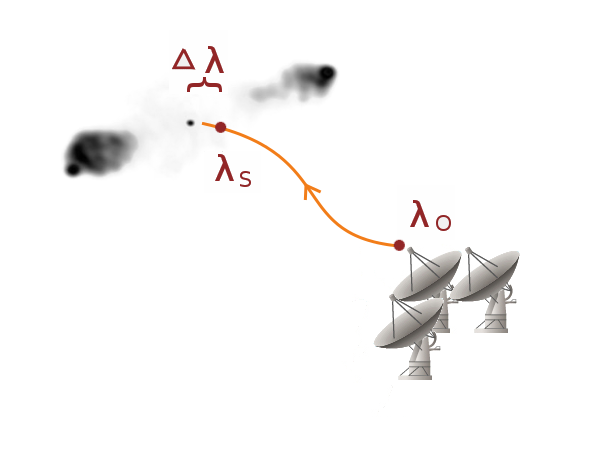}
 \caption{Affine parameter convention of a light ray in a radio observation.}
 \label{fig:affine}
\end{figure}

The authors of ~\cite{Sasaki1987,Pyne1993} 
suggest that the analysis of perturbed null geodesics is drastically simplified 
by means of a conformal transformation. Below we follow this approach and regard the cosmic scale factor 
to be a conformal transformation of a perturbed Minkowski space-time. 
The redshift is then defined as
\begin{equation}
z = \frac{\omega_{\rm s}}{\omega_{\rm o}} - 1 =
\frac{a_{\rm o}(u_\mu k^\mu)_{\rm s}}{a_{\rm s}(u_\nu k^\nu)_{\rm o}} - 1,
\end{equation}
where $\omega_{\rm s}$ and $\omega_{\rm o}$ are the frequencies at the source and observer, 
respectively. In our notation and at linear order 
\begin{eqnarray}
1+z&=&\frac{\mathcal{A}_{\rm o}}{\mathcal{A}_{\rm s}}[1-\Phi|^{\rm s}_{\rm o}+e^{ri}
V_i|^{\rm s}_{\rm o}+k^0\int^{\lambda_{\rm s}}_{\lambda_{\rm o}}{\rm d}\lambda'(\dot{\Phi}-\dot{\Psi})
\nonumber\\
&&-\frac{1}{2}k^0\int^{\lambda_{\rm s}}_{\lambda_{\rm o}}
{\rm d}\lambda'e^{ri}e^{rj}(U_{i,j}+U_{j,i}+\dot{h}_{ij}) ] , \label{equ:rdg}
\end{eqnarray}
with the gauge invariant ratio of scale factors 
$\frac{\mathcal{A}_{\rm o}}{\mathcal{A}_{\rm s}}\equiv\frac{a_{\rm o}}{a_{\rm s}}(1-\mathcal{H}(B+\dot{E})|^{\rm s}_{\rm o})$ and the gauge invariant velocity $V_i\equiv v_i-S_i+\dot{E}_{,i}$. Thus this expression is manifestly gauge invariant.
The  affine parameter $\lambda$ is related to conformal time via ${\rm d} \eta = k^0 {\rm d} \lambda$, see 
Eq.~(\ref{A2}). Our sign convention and the notation is illustrated in Fig.~\ref{fig:affine}.

For convenience, we define  $1+z \equiv \frac{a_{\rm o}}{a_{\rm s}}(1+\delta z)$, 
and thus the redshift distortion becomes
\begin{eqnarray}
\delta z&=&-\mathcal{H}(B+\dot{E})|^{\rm s}_{\rm o}-\Phi|^{\rm s}_{\rm o}+e^{ri}
V_i|^{\rm s}_{\rm o}+k^0\int^{\lambda_{\rm s}}_{\lambda_{\rm o}}{\rm d}\lambda'(\dot{\Phi}-\dot{\Psi})
\nonumber\\
&&-\frac{1}{2}k^0\int^{\lambda_{\rm s}}_{\lambda_{\rm o}}
{\rm d}\lambda'e^{ri}e^{rj}(U_{i,j}+U_{j,i}+\dot{h}_{ij}) .
\end{eqnarray}
More details are provided in Appendix \ref{App:B}. From this equation, one could clearly see the well-known gravitational redshift, the Doppler shift and the integrated Sachs-Wolfe effect, as well as vector mode 
and gravitational wave contributions. 
We also introduce the notation $1+\bar{z}\equiv\frac{a_{\rm o}}{a_{\rm s}}$ to indicate the unperturbed redshift, which will be used later.

\subsection{ Specific flux fluctuations}

Since radio sources typically have featureless (power law) spectra, their redshift cannot be 
obtained from radio continuum observations. However, we observe the specific flux. The  
observed specific flux of a radio source is also affected by metric fluctuations.
This effect modifies any distance estimate based on the ratio of specific fluxes (assuming for a moment 
that we would know the specific luminosities).

The energy momentum tensor of a radio source is
\begin{equation}
 T^{\mu\nu}= \frac{1}{8\pi}\int {\rm d\omega}  \mathscr{A}^2(\omega,\lambda) \hat{k}^\mu \hat{k}^\nu.
\end{equation}
Here, we use $\hat{k}^\mu$ to distinguish the physical wave vector from $k^\mu$, 
the wave vector in the conformally related Minkowski space-time.
The bolometric flux is given by a projection of the energy-momentum,
\begin{equation}
S_{\rm bol} \equiv - e_\alpha u_{\rm o}^\nu T^\mu_\nu h^\alpha_\mu,  
\end{equation}
where $h^\alpha_\mu$ is the spatial projection tensor, orthogonal to the observer
four-velocity $u^\nu_{\rm o}$, and $e_\alpha$ is a unit space like
vector pointing in the direction of the 3 wave vector in the observer rest frame.
These vectors are defined at the observer, and we parallel
transport the wave vector and energy-momentum tensor along the geodesic.
Since light rays with different frequency follow the same geodesic, we find
\begin{equation}
S_{\rm bol} =\frac{1}{8\pi}\int {\rm d\omega} \mathscr{A}^2(\omega, \lambda)\omega^2.
\end{equation}
Therefore, the specific flux density is
\begin{equation}
S(\omega) =\frac{1}{8\pi}\mathscr{A}^2(\omega,\lambda)\omega^2.
\end{equation}
At long wavelengths, synchrotron radiation is the dominant radiation process, which suggests that 
the emitted flux density follows a power law, 
\begin{equation}
S_{\rm s}(\omega_{\rm s})\propto \omega^{-\alpha}_{\rm s}, 
\end{equation}
where $\alpha$ is the spectral index.

The emitted photon number in a frequency band of width ${\rm d} \omega_{\rm s}$, 
solid angle  ${\rm d}\Omega_{\rm s}$, and proper time interval ${\rm d}\tau_{\rm s}$ 
can be expressed in terms of the specific luminosity of a source 
$L(\omega_{\rm s}) \equiv 
4\pi R_{\rm s}^2 S_{\rm s} (\omega_{\rm s})$ ($R_{\rm s}$ is a distance not too far from 
the center of the source) and reads
\begin{equation}
{\rm d N_\gamma =  \frac{L(\omega_{\rm s})}{4\pi \omega_{\rm s}} 
{\rm d}\omega_{\rm s}{\rm d}\Omega_{\rm s} {\rm d}\tau_{\rm s}.} 
\end{equation}
Due to the conservation of photon number (neglecting absorption and emission along the line of sight 
to a source) we can relate that to the observed specific flux density ~\cite{Schneider},
\begin{eqnarray}
 \frac{L(\omega_{\rm s})}{4\pi \omega_{\rm s} } {\rm d}\omega_{\rm s}{\rm d}\Omega_{\rm s} {\rm d}\tau_{\rm s}
=    \frac{S_{\rm o}(\omega_{\rm o})}{\omega_{\rm o}}{\rm d}\omega_{\rm o}{\rm d}A_{\rm o}{\rm d}\tau_{\rm o}.   
\end{eqnarray}
The (monochromatic) luminosity distance $D_{\rm L}$ is
\begin{equation}\label{eq:fluxratio}
D_L\equiv \sqrt{\frac{L_{\rm s}(\omega_{\rm s}) {\rm d}\omega_{\rm s}}{ 4\pi S_{\rm o}(\omega_{\rm o}){\rm d}\omega_{\rm o}}} =R_{\rm o}(1+z),
\end{equation}
where we introduce the physical distance (today) $R_{\rm o} \equiv 
\sqrt{{\rm d}A_{\rm o}/{\rm d}\Omega_{\rm s}}$.
$D_L$ agrees with the luminosity distance inferred from the bolometric flux of a thermal source. 

To infer the distance of a source that is neither monochromatic 
nor thermal requires the detailed knowledge of its spectrum (besides its luminosity).
For featureless spectra the redshift is typically unknown. 
It is thus convenient to compare the observed specific flux density to the specific luminosity 
at the observed frequency and we use the observed bandwidth. We define the specific luminosity 
distance,
\begin{equation}\label{eq:sld}
D_S \equiv \sqrt{\frac{L_{\rm s}(\omega_{\rm o}) {\rm d}\omega_{\rm o} }
{4\pi S_{\rm o}(\omega_{\rm o}) {\rm d}\omega_{\rm o}}} 
= (1+z)^{(\alpha-1)/2} D_L.
\end{equation}

The last term in Eq.~(\ref{eq:sld}) connects the specific luminosity distance with the 
(monochromatic/bolometric) luminosity distance $D_{\rm L}$, the latter was discussed 
many times ~\cite{Sasaki1987,Bonvin2006,Hui2006,Yoo2009a}.  

\begin{figure}
\centering
 \includegraphics[width=0.8\linewidth]{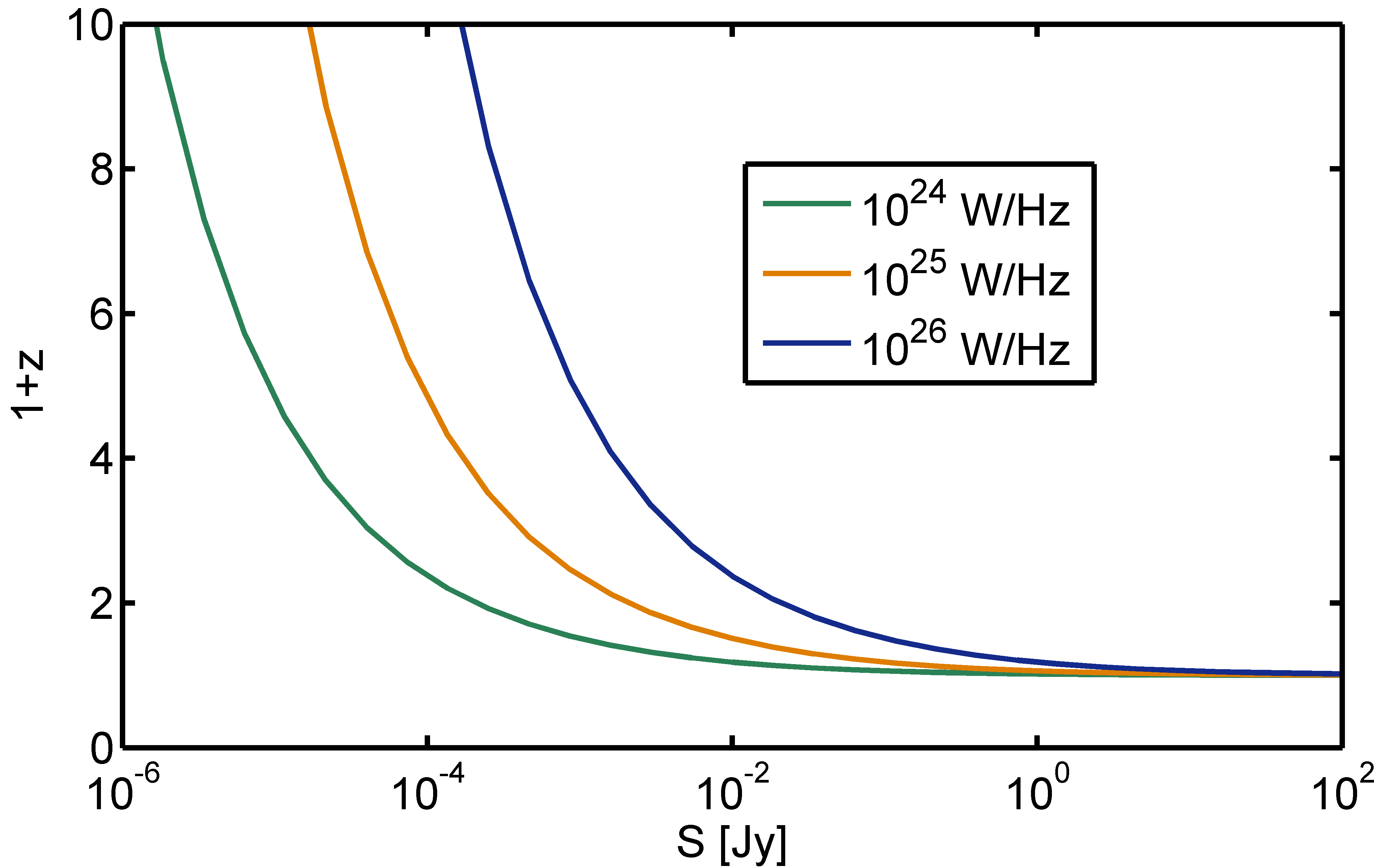}
 \caption{Flux-redshift relation for different specific luminosities typical for AGNs. The standard $\rm{\Lambda}$ cold dark matter model has been adopted, and the spectral index $\alpha$ has been chosen to be $0.75$. }
 \label{fig:redshift-flux}
\end{figure}

In the following our task is to calculate the specific flux density of a radio source, taking all 
linear fluctuations into account. We can write
\begin{equation}
S_{\rm o} (\omega_{\rm o}) = S_{\rm s}(\omega_{\rm o}) \frac{R_{\rm s}^2}{D_S^2} = 
\frac{L_{\rm s}(\omega_{\rm o})}{4 \pi} \frac{1}{(1+z)^{\alpha + 1} R_{\rm o}^2}. 
\end{equation}
For the standard cosmological (homogeneous and isotropic) model, this relation between flux 
density and redshift is shown in Fig.~\ref{fig:redshift-flux} for several typical specific luminosities 
of radio sources. 
The linear distortions of redshift were presented in the previous subsection. Analogous to the 
redshift distortion, we define the physical distance fluctuation $\delta_d$ via
\begin{equation}
R_{\rm o} =  \frac{D_{\rm L}}{1+z} = a_{\rm o}(\eta_{\rm o}-\eta_{\rm s})[1+\delta_d],
\end{equation}
where $\delta_d$ is then given by comparison with the expression for the luminosity 
distance at linear order, which has been discussed previously ~\cite{Sasaki1987,Bonvin2006,Hui2006}. 

As shown in the appendix, $R_{\rm o}$ can be expressed in terms of gauge
invariant quantities as 
\begin{widetext}
\begin{eqnarray}\label{eq:lum}
R_{\rm o} &=&a_{\rm o}(\eta_{\rm o}-\eta_{\rm s}) \bigg[1+\psi_{\rm o}-\Phi_{\rm s}+\Psi_{\rm s} +e^i
(V_i)_{\rm s}+\frac{[\dot{E}+B]|^{\rm s}_{\rm o}}{\eta_{\rm s}-\eta_{\rm o}}
+k^0\int^{\lambda_{\rm s}}_{\lambda_{\rm o}}(\dot{\Phi}-\dot{\Psi}){\rm d}\lambda+\frac{2}{
\eta_{\rm s}-\eta_{\rm o}}\int^{\lambda_{\rm s}}_{\lambda_{\rm o}}k^0(\Phi-\Psi) {\rm d}\lambda  \nonumber\\
&&
-\frac{1}{\eta_{\rm s}-\eta_{\rm o}}\int^{\lambda_{\rm s}}_{\lambda_{\rm o}} 
(\lambda_{\rm s}-\lambda)k^2\big(\dot{\Phi}-\dot{\Psi} \big){\rm d}\lambda 
-\int^{\lambda_{\rm s}}_{\lambda_{\rm o}} {\rm d}\lambda
\frac{(\lambda_{\rm s}-\lambda)(\lambda-\lambda_{\rm o})}{2(\lambda_{\rm s}-\lambda_{\rm o})}k^2\bigg[
\Delta(\Phi-\Psi)-(\Phi-\Psi)_{,ij}e^ie^j\bigg]  \nonumber\\
&&-k^0\int^{\lambda_{\rm s}}_{\lambda_{\rm o}}\frac{1}{2}e^ie^j(U_{i,j}+U_{j,i}+\dot{h}_{ij}
) {\rm d}\lambda
+\frac{1}{\eta_{\rm s}-\eta_{\rm o}}\int^{\lambda_{\rm s}}_{\lambda_{\rm o}} 
(\lambda_{\rm s}-\lambda)k^2\big[\frac{1}{2}e^ie^j(U_{i,j}+U_{j,i}+\dot{h}_{ij})\big]
{\rm d}\lambda  \nonumber\\
&&-\int^{\lambda_{\rm s}}_{\lambda_{\rm o}} {\rm d}\lambda
\frac{(\lambda_{\rm s}-\lambda)(\lambda-\lambda_{\rm o})}{2(\lambda_{\rm s}-\lambda_{\rm o})}k^2\big[
\frac{1}{2}(\dot{U}_{i,j}+\dot{U}_{j,i}+\ddot{h}_{ij}-\Delta
h_{ij})e^ie^j-\Delta U_ie^i\big]\bigg].
\end{eqnarray}
 \end{widetext}
We have checked that $R_{\rm o}$ is manifestly gauge invariant, and after gauge 
fixing our result, it agrees with Bonvin et al.~\cite{Bonvin2006}.  

As shown so far, distortions of the specific flux are affected by redshift 
distortions $\delta z$ and physical distance fluctuations $\delta_d$. Besides these geometrical effects, 
the specific luminosity and spectra of different sources are not identical, which provides another source of fluctuation.
Thus,  we allow $L_{\rm s}(\omega_{\rm o})$ and $\alpha$ to vary and denote its fluctuations by $\delta L_{\rm s}(\omega_{\rm o})=L_{\rm s}(\omega_{\rm o})-\bar{L_{\rm s}}(\omega_{\rm o})$ and $\delta \alpha = \alpha - \bar \alpha$, respectively. 
The specific flux density can be written as 
\begin{equation}
 S_{\rm o}(\omega_{\rm o}) =\bar{S}_{\rm o}(\omega_{\rm o})(1+\delta_S),
\end{equation}
where
\begin{equation}\label{eq:bkS}
\bar{S}_{\rm o}(\omega_{\rm o})=\frac{\bar{L}_{\rm s}(\omega_{\rm o})}{4\pi a_{\rm o}^2 (1+\bar{z})^{\bar{\alpha}+1}(\eta_{\rm o}-\eta_{\rm s})^2}
\end{equation}
and the specific flux fluctuation is
\begin{equation}
 \delta_S =\frac{\delta L_{\rm s}(\omega_{\rm o})}{\bar{L}_{\rm s}(\omega_{\rm o})} - 2\delta_d-(\bar{\alpha}+1)\delta z-\delta \alpha \ln(1+ \bar{z}).
\end{equation}
On one hand, at high redshifts and large fields of view (a large sample) the geometric terms $(- 2\delta_d-(\bar{\alpha}+1)\delta z)$ are likely to 
dominate $\delta_S$. On the other hand, at low redshift and small fields of view, $\delta \alpha$ and $\delta L_{\rm s}$ may play a significant role,
 which might explain some of the variation observed in the differential number counts in small fields.

\subsection{Number counts in observed spherical  coordinates and lensing effect}

\begin{figure}
\centering
 \includegraphics[width=0.8\linewidth]{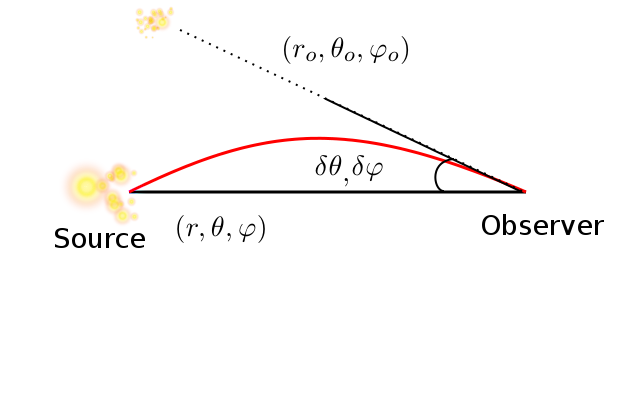}
 \caption{Observed position vs. background position.}
 \label{fig:geodesic_new}
\end{figure}

As a result of the fluctuations we mentioned above, we have to taken them into account when we do the coordinate transformation from the background coordinates $(r,\theta,\varphi)$ to the observed coordinates 
$(r_{\rm o},\theta_{\rm o},\varphi_{\rm o})$ (see Fig.~\ref{fig:geodesic_new}). 
We assume that the two sets of coordinates are related 
by small quantities, such that
\begin{eqnarray}
&& r = r_{\rm o} + \delta r, \nonumber \\
&& \theta=\theta_{\rm o}+\delta\theta,  \\
&& \varphi = \varphi_{\rm o}+\delta\varphi.  \nonumber
\end{eqnarray} 
The comoving distance fluctuation is defined as the difference between the line of sight distance $r$ in 
the comoving coordinates and the distance $r_{\rm o}$ inferred from the observed flux density $S_{\rm o}$ for a fixed luminosity, 
measured spectral index and assumed luminosity. Unlike the former, 
$r_o$ is in principle a measurable quantity and it is invariant under coordinate transformations.

The observed flux is a function of the conformal time. 
Using $\bar{r}\equiv\eta_{\rm o}-\eta$ and Eq.~(\ref{eq:bkS}), we explicitly define the function $\bar{r}=\bar{r}(\bar{S}_{\rm o})$,
 and the inferred distance $r_{\rm o} \equiv \bar{r}(S_{\rm o})$.  Expanding this definition at background flux leads to
\begin{eqnarray}
\bar{r}(S_{\rm o})&=&\bar{r}(\bar{S}_{\rm o})+\frac{{\rm d}\bar{r} }{{\rm d}\bar{S}_{\rm o}}(S_{\rm o}-\bar{S}_{\rm o}) \nonumber\\
 r_{\rm o}&=&\eta_{\rm o}-\eta- \frac{ (\eta_{\rm o}-\eta)\delta_{S}}{2+(\bar{\alpha}+1)(\eta_{\rm o}-\eta)\mathcal{H}}.
\end{eqnarray}
The linear order light cone relation in the background coordinates is
\begin{equation}
\eta_{\rm o}-\eta-\int_{\lambda}^{\lambda_{\rm o}} {\rm d}\lambda' l^0= r-\int^{\lambda}_{\lambda_{\rm o}} {\rm d}\lambda
l^i e^{r}_i,
\end{equation}
where $l^\mu$ is the wave vector fluctuation caused by metric fluctuation in the conformally 
related geometry, for more details see Appendix \ref{App:B}. 
According to the null condition Eq.~(\ref{eq:nulk}), one can find the inferred distance deviation
\begin{eqnarray}
\delta r&=& r-r_{\rm o} \\
&=& \frac{ r_{\rm o}\delta_{S}}{2+(\bar{\alpha}+1)r_{\rm o}\mathcal{H}}+[\frac{k^iF_i}{k^0}-B+\frac{k^iE_{,i}}{k^0}-\dot{E
}  ]|^{\rm s}_{\rm o}
\nonumber\\
&&+\int_{\rm o}^{\rm s} {\rm d}\lambda (-k^0\Phi+k^0\Psi- U_i k^i +\frac{k^ik^j}{2k^0}h_{ij}) ,\nonumber
\end{eqnarray}
where we have replaced $\eta_{\rm}-\eta$ by $r_{\rm o}$, which introduces contributions at higher order that we neglect.
Metric perturbations can deflect and disperse light rays and thus displace 
the observed angles on the sky (see Fig.~\ref{fig:geodesic_new}). 
Following ~\cite{Bonvin2011}, 
\begin{eqnarray}\label{eq:len1}
\delta \theta &=&-\frac{1}{r_{\rm o}}\int^{\rm s}_{\rm o} {\rm d}\lambda \bigg( e^{\theta i}[g^{(1)}_{0
i}k^0+g^{(1)}_{j i}k^j]^{\rm s}_{\rm o}\nonumber\\
&&-  \frac{\lambda_{\rm s}-\lambda}{2r} g^{(1)}_{\sigma \rho,\theta}k^{\sigma}
k^{\rho} \bigg) ,
\end{eqnarray}
\begin{eqnarray}\label{eq:len2}
\delta \varphi &=&-\frac{1}{r_{\rm o} \text{sin}(\theta_{\rm o})}\int^{\rm s}_{\rm o} {\rm d}\lambda \bigg(
e^{\varphi i}[g^{(1)}_{0 i}k^0+g^{(1)}_{j i}k^j]^{\rm s}_{\rm o} \nonumber\\
&&- \frac{\lambda_{\rm s}-\lambda}{2r \text{sin}(\theta_{\rm o})} g^{(1)}_{\sigma
\rho,\varphi}k^{\sigma} k^{\rho}  \bigg) ,
\end{eqnarray}
where $e^{\theta i}$ and $e^{\varphi i}$ are the unit vectors point into the angular direction. 
For further details see Appendix \ref{App:An} and ~\cite{Bonvin2011}.

The Jacobian of the transformation from the background coordinates to observed
coordinates  is 
\begin{equation}
\det(J)=\frac{1}{S_o}\frac{ -r_{\rm o}}{2+(\bar{\alpha}+1)r_{\rm o}\mathcal{H}} [1+\frac{{\rm d} \delta r}{{\rm d} r_{\rm o}}+\frac{\partial
\delta \theta}{\partial \theta_{\rm o}}+\frac{\partial \delta \varphi}{\partial
\varphi_{\rm o}}].
\end{equation}
The prefactor is gauge invariant, since it is the derivative of observed flux respect to the inferred distance $r_{\rm o}$.
To the linear order,
\begin{equation}
\frac{{\rm d} \delta r}{{\rm d} r_{\rm o}}=\frac{\partial \delta r}{\partial r}-\frac{\partial\delta r}{\partial \eta} ,
\end{equation}
and according to the transformation law of vectors,
the three-velocity of the source can be expressed in the observed coordinates as
\begin{equation}
 V'^i=v^i-\frac{\partial \delta x^i}{\partial \eta}.
\end{equation}

We combine the previous results with the expression for Eq.~(\ref{eq:covolume}) to obtain the total number count for the sources with identical luminosity.
\begin{eqnarray}\label{eq:result1}
N&=&\int {\rm d}\Omega_{\rm o} \int \frac{{\rm d}S_{\rm o}}{S_{\rm o}}   \frac{a^3 r_{\rm o}^3 n_{\rm phy}}{2+(\bar{\alpha}+1)r_{\rm o}\mathcal{H}} [1+3\psi+\Delta E \nonumber \\
&&+V'^ie_i^r +2\frac{\delta r}{r_{\rm o}}+\frac{\partial \delta
r}{\partial r_{\rm o}}-2\kappa_{\rm g}],
\end{eqnarray}
where,  we changed $\partial \delta r / \partial r$ to $\partial \delta r / \partial r_{\rm o}$, since $\delta r$ is a first order quantity.
$\kappa_{\rm g}$ denotes the gravitational lensing convergence,
\begin{eqnarray}\label{eq:kappa-d}
 \kappa_{\rm g}&=&-\frac{1}{2}[(\cot \theta_{\rm o}+\frac{\partial}{\partial \theta_{\rm o}})\delta
\theta+\frac{\partial \delta \varphi}{\partial \varphi_{\rm o}}].
 \end{eqnarray}
Inserting the angular displacements into Eq.~(\ref{eq:kappa-d}), we get 
\begin{widetext}
\begin{eqnarray}\label{eq:kappa}
&&\kappa_{\rm g}= \frac{1}{2r_{\rm o}}\int^{\rm s}_{\rm o} {\rm d}\lambda (\lambda_{\rm s}-\lambda)\left[\frac{1}{2  r(\lambda)}
\hat{\nabla}^2\left(2k^2(\Phi-\Psi)+ 2U_i k^i k^0 -k^ik^jh_{ij}\right)\right. \\
&& \left. +[(\cot \theta_{\rm o} + \frac{\partial}{\partial \theta_{\rm o}} )e^{\theta
i}  +\frac{1}{\text{sin}(\theta_{\rm o})}\frac{\partial}{\partial \varphi_{\rm o}}
e^{\varphi i}  ]
\frac{{\rm d}}{{\rm d} \lambda}(-U_i  k^0 +h_{ij}k^j ) 
+[(\cot \theta_{\rm o} + \frac{\partial}{\partial \theta_{\rm o}} )e^{\theta
i}  +\frac{1}{\sin(\theta_{\rm o})}\frac{\partial}{\partial \varphi_{\rm o}}
e^{\varphi i}  ]
\frac{{\rm d}^2}{{\rm d} \lambda^2}\left( E_{,i} + F_i \right) \right], \nonumber
\end{eqnarray}
\end{widetext}
where $\hat{\nabla}^2$ is the Laplacian operator on a unit sphere,
\begin{equation}
 \hat{\nabla}^2=\cot \theta_{\rm o} \frac{\partial}{\partial \theta }  +
\frac{\partial^2}{\partial \theta_{\rm o}^2}+ \frac{1}{ \sin^2(\theta_{\rm o}) }
\frac{\partial^2}{\partial \varphi_{\rm o}^2}.
\end{equation}

According to its definition, $\kappa_{\rm g}$ describes the solid angle difference between the observer coordinates and the background coordinates.
Since the background coordinates are not measurable, $\kappa_{\rm g}$ changes under coordinate transformations.
A gauge invariant quantity
\begin{eqnarray}
K_{\rm g}&\equiv& \kappa_{\rm g}-\frac{1}{2r_{\rm o}}\int^{\rm s}_{\rm o} {\rm d}\lambda (\lambda_{\rm s}-\lambda) \big[(\cot \theta_{\rm o} + \frac{\partial}{\partial \theta_{\rm o}} )e^{\theta i} \\  
&&+\frac{1}{\sin(\theta_{\rm o})}\frac{\partial}{\partial \varphi_{\rm o}}
e^{\varphi i}  \big]\frac{{\rm d}^2}{{\rm d} \lambda^2}\left( E_{,i} + F_i \right)  \nonumber
\end{eqnarray}
can be inferred from the angular diameter distance fluctuations. 
After gauge fixing, $K_{\rm g}$ agrees with the gravitational lensing convergence in \cite{2014MNRAS.443.1900B,2008PhRvD..78l3530B}. 
Additionally, it is useful to also define a gauge invariant comoving distance fluctuation
\begin{equation}
 \delta R\equiv \delta r-[\frac{k^iF_i}{k^0}+\frac{k^iE_{,i}}{k^0}  ]|^{\rm s}_{\rm o}.
\end{equation}
The gauge dependent contributions in $\kappa_{\rm g}$ and $\delta r$, that depend on
the position of the source, cancel the $\Delta E$ term (from $\sqrt{-g}$)
in Eq.~(\ref{eq:result1}).

We have checked that the result Eq.~(\ref{eq:result1}) agrees with ~\cite{Yoo2010} at the background level after 
replacing flux density by redshift. Our comoving distance fluctuation is different due to the different choice of observable( we consider flux density, they consider redshift).

\subsection{Physical number density}

In the previous sections, we have evaluated the effect of metric fluctuations on the total number count for a fixed luminosity.
The next step is to integrate all possible source luminosities.
We assume
\begin{eqnarray}
&&n_{\rm phy}(L,S_{\rm o},\theta_{\rm o},\varphi_{\rm o})=\sum_i n_i \\
&=&\bigg(\frac{a_o}{a}\bigg)^3 \sum_i \rho_i(L) p_i(L,r_{\rm o})(1+\delta_{n_i}(L,S_{\rm o},\theta_{\rm o},\varphi_{\rm o}))\nonumber ,
\end{eqnarray}
where the index $i$ characterizes the different types of sources (e.g.~AGN or SFG, or any finer classification),  
$\rho_i(L)$ and $p_i(L,r_{\rm o})$ are the local (today's) luminosity function and the generalized evolution
function ~\cite{Longair}.
It is common to parametrize the luminosity function in terms of a 
two-power law function ~\cite{Dunlop1990},
\begin{eqnarray}
 \rho(L)=\rho_{c}[(\frac{L}{L_{c}})^\beta +(\frac{L}{L_{c}})^\gamma]^{-1}. 
\end{eqnarray}
Other prominent functions are the Schechter luminosity function ~\cite{Schechter1976}
\begin{eqnarray}
 \rho(L)=\rho_{c}(\frac{L}{L_{c}})^{-\beta}\exp(-\frac{L}{L_{c}}), 
\end{eqnarray}
or a simple power-law.

Here we also introduce gauge invariant number density perturbation $\Delta_{n_i}$,
\begin{equation}
 \Delta_{n_i}=\delta_{n_i}+3\psi
\end{equation}

\section{\label{sec:4} Differential number counts}

In the early days of cosmology, integral number counts $N(\geq S)$ have been used 
quite frequently. However, this is not the best way to represent the data, as error evaluation for 
such a cumulative quantity is sophisticated. Alternatively, the differential number counts, 
i.e.~the number of sources inside the flux interval $S$ to $S+\Delta S$, are a good
alternative. 

We thus arrive at the central result of this work, the expression for the differential number counts 
including all linear order effects:

\begin{eqnarray}
&&\frac{ {\rm d}^2 N }{ {\rm d \, ln} S_{\rm o} {\rm d} \Omega_{\rm o} }  \\
&=&-\sum_{i}\int^\infty_0 {\rm d}L \rho_i(L) p_i(L,r_{\rm o}) \frac{a_o^3 r_o^3 }{2+(\bar{\alpha}_i+1)r_o\mathcal{H}} \times \nonumber \\
&&[1+\Delta_{n_i}+V'\cdot e^r +2\frac{\delta R}{r_o}+\frac{\partial \delta
R}{\partial r_o}-2K_{\rm g}].\nonumber 
\end{eqnarray}

\section{Discussion and conclusion}

We present a theoretical framework for the prediction of differential number counts, either analytically or 
by means of simulations. This framework is based on fully relativistic linear perturbations of a spatially 
flat, isotropic and homogeneous  space-time metric. In particular we did not assume any gauge condition. 
We have checked that the number of sources within fixed intervals of flux, frequency and solid angle 
is gauge invariant.

In previous work ~\cite{Bonvin2011,Challinor2011,Yoo2009}, the number density 
has been studied as a function of redshift. There the redshift distortion is one of the dominant effects in 
the radial direction. In our case, as shown in the Sec. \ref{sec:OB} the radial direction 
fluctuation comes from three effects, i.e.~redshift distortions, physical distance fluctuations, 
variation of the source luminosities and spectral indices. This makes the evaluation more involved 
than the case when the redshifts of each source are accessible.

With the complete derivations of the covariant volume integral on the past light
cone, we have identified several contributions in the differential number count fluctuations, including
Doppler effect, generalized Sachs-Wolfe effects, lensing effect and astrophysical variations 
(luminosity and spectral index).

To further constrain the differential number counts will require
not only accurate theoretical predictions, but also to model and measure luminosity functions, 
luminosity and density evolution. However, the luminosity and density evolution of galaxies is not 
important when we study the statistical properties of $n$-point correlations on large enough scales.  
A more detailed analysis, especially in the light of planned radio surveys with ASKAP, MeerKAT, 
LOFAR and SKA will be presented elsewhere.

\appendix
\section{\label{App:B} Null geodesics and redshift}

Conformal transformations preserve the causal structure of space-time. 
Thus we can find the null geodesics of a linearly perturbed Minkowski space-time and 
relate them to the null-geodesics of the spatially flat Friedmann-Lema\^itre cosmologies 
via a conformal transformation provided by the scale factor. This strategy was used in 
~\cite{Sasaki1987,Pyne1993,Bonvin2006}. For completeness, we repeat the 
most essential steps in our notation.

The null geodesic $x^\mu(\lambda)$, with $\lambda$ denoting an affine parameter,
can be decomposed into a background path plus a perturbation,
\begin{equation}
x^\mu(\lambda)=x^{(0) \mu}(\lambda)+ x^{(1) \mu}(\lambda),
\end{equation}
where $x^{(0) \mu}$ is a null geodesic in Minkowski space-time, 
and we assume that the metric perturbations are small. 
The null vector-field is therefore
\begin{equation}
\label{A2}
 k^\mu=\frac{{\rm d} x^{(0)\mu}}{{\rm d} \lambda}, \;\;l^{ \mu}=\frac{{\rm d} x^{(1)
\mu}}{{\rm d}\lambda} .
\end{equation}
At the first order, the null condition becomes
\begin{eqnarray}\label{eq:nulk}
 & & -k^0 l^0+k^i l^i = k^2\phi +(B_{,i}+S_i) k^i k^0 \nonumber\\
 & & \qquad - k^ik^j(\psi \delta_{ij} 
 +E_{,ij} +\frac{1}{2}F_{i,j}+\frac{1}{2}F_{j,i}+\frac{1}{2}h_{ij}), 
\end{eqnarray}
where we define $(k^0)^2=(k^ie_i)^2\equiv k^2$.

Now we turn to the perturbed geodesic equation. The zeroth order geodesic equation 
simply tells us that $x^{(0)\mu}$ is a straight trajectory, while the first order geodesic equation is
\begin{equation}
 \frac{{\rm d} l^\mu}{{\rm d} \lambda}=-2\Gamma^{(0)\mu}_{\rho\sigma}k^\rho l^\sigma-
\Gamma^{(1)\mu}_{\rho\sigma}k^\rho
k^\sigma-\Gamma^{(0)\mu}_{\rho\sigma,\nu}k^\rho k^\sigma x^{(1)\nu}.
\end{equation}
For the flat background,
\begin{equation}\label{eq:delta}
 \frac{{\rm d} l^\mu}{{\rm d} \lambda}=- \Gamma^{(1)\mu}_{\rho\sigma}k^\rho k^\sigma.
\end{equation}
The temporal component of this equation is
\begin{eqnarray}\label{equ:l0}
\frac{{\rm d} l^0}{{\rm d} \lambda}
&=&-2\frac{{\rm d} \phi}{{\rm d} \lambda}k^0+k^2[\dot{\phi}-\dot{\psi}]-k^ik^j [\dot{E}_{,ij}+B_{,ij} \nonumber\\
&& +\frac{1}{2}(S_{i,j}+S_{j,i}+\dot{F}_{i,j}+\dot{F}_{j,i})+\frac{1}{2}\dot{h}_{
ij}],
\end{eqnarray}
where we used ${\rm d \phi}/{\rm d}\lambda = \dot{\phi} {\rm d}\eta/{\rm d}\lambda + 
\phi_{,i}{\rm d} x^i/{\rm d}\lambda$.

After integrating Eq.~(\ref{equ:l0}), we obtain the temporal component of the wave number
perturbation
\begin{eqnarray}
l^0|^{\rm s}_{\rm o} &=&
- 2 k^0 \phi|^{\rm s}_{\rm o} + 
  k^2 \int^{\lambda_{\rm s}}_{\lambda_{\rm o}} {\rm d} \lambda' [\dot{\phi}-\dot{\psi}]
- k^2 \int^{\lambda_{\rm s}}_{\lambda_{\rm o}} {\rm d} \lambda'e^{ri}e^{rj}[\dot{E}_{,ij}\nonumber \\   
&&+B_{,ij}+\frac{1}{2}(S_{i,j}+S_{j,i}+\dot{F}_{i,j}+\dot{F}_{j,i}+\dot{h}_{ij})],   
\end{eqnarray}
where $e^{ri}$ denotes the unit vector pointing from the observer to the source.
With $l^0$ one can further evaluate the redshift at linear order, by means of
\begin{equation}
 1+z =\frac{a_{\rm o}}{a_{\rm s}}\;
 \frac{(k\cdot u^{(0)})_{\rm s}+(k\cdot u^{(1)})_{\rm s}+(l\cdot u^{(0)})_{\rm s}}
 {(k\cdot u^{(0)})_{\rm o}+(k\cdot u^{(1)})_{\rm o}+(l\cdot u^{(0)})_{\rm o}}.
\end{equation}
Since $u^i$ is of first order, only the time component of $l^\mu$ contributes and we find
\begin{eqnarray}
1+z&=&\frac{A_{\rm o}}{A_{\rm s}}[1-\Phi|^{\rm s}_{\rm o}+e^{ri}
V_i|^{\rm s}_{\rm o}+k^0\int^{\lambda_{\rm s}}_{\lambda_{\rm o}}{\rm d}\lambda'(\dot{\Phi}-\dot{\Psi})
\nonumber\\
&&-\frac{1}{2}k^0\int^{\lambda_{\rm s}}_{\lambda_{\rm o}}
{\rm d}\lambda'e^{ri}e^{rj}(U_{i,j}+U_{j,i}+\dot{h}_{ij}) ]. 
\end{eqnarray}

\section{\label{App:An} Angular displacement}

Equations~(\ref{eq:len1}) and (\ref{eq:len2}) were derived in ~\cite{Bonvin2011}. Therefore, we 
just provide the most essential steps. 

We start from an infinitesimal deviation in the $\hat{\theta}$ direction,
\begin{equation}
 r \delta \theta=e_{\theta i}\delta x^i= \int^{\rm s}_{\rm o} {\rm d} \lambda e_{\theta i} l^i.
\end{equation}
Since angles are not affected by conformal transformations, $\delta \theta$ can
be calculated from the geodesic equation in the conformally related geometry,
\begin{eqnarray}
 \frac{{\rm d}  l^i }{{\rm d} \lambda}&=& - \Gamma^{(1) i}_{\sigma\rho} k^{\sigma} k^{\rho}
\nonumber\\
 &=&-\frac{1}{2}\delta^{i\alpha} ( g^{(1)}_{\sigma \alpha,\rho}+g^{(1)}_{\rho
\alpha,\sigma}-g^{(1)}_{\sigma \rho,\alpha})k^{\sigma} k^{\rho} \nonumber\\
 &=& -\frac{{\rm d} \delta^{i\alpha} g^{(1)}_{\sigma \alpha} }{{\rm d} \lambda} k^{\sigma}
+\frac{1}{2}\delta^{i\alpha}g^{(1)}_{\sigma \rho,\alpha}k^{\sigma} k^{\rho} .
\end{eqnarray}
With the help of $\frac{{\rm d} k^\mu}{{\rm d} \lambda}=\Gamma^{(0)\mu}_{\sigma
\rho} k^{\sigma} k^{\rho}=0$, we find 
\begin{equation}
 e_{\theta i} l^i |^{\rm s}_{\rm o}=-e^{\theta i}[g^{(1)}_{0 i}k^0+g^{(1)}_{j i}k^j]^{\rm s}_{\rm o}
+\frac{1}{2}\int^{\rm s}_{\rm o} {\rm d} \lambda e^{\theta i} g^{(1)}_{\sigma \rho,i}k^{\sigma}
k^{\rho} \nonumber.
\end{equation}
Integrate $e_{\theta i} l^i$ along the path to obtain 
\begin{eqnarray}
\delta \theta &=&
-\frac{1}{r_{\rm o}}\int^{\rm s}_{\rm o} {\rm d}\lambda 
\bigg(e^{\theta i}[g^{(1)}_{0i}k^0 + g^{(1)}_{j i}k^j]^{\rm s}_{\rm o} \nonumber \\
&&
-\frac{\lambda_{\rm s}-\lambda}{2r} g^{(1)}_{\sigma \rho,\theta}k^{\sigma}
k^{\rho} \bigg),  
\end{eqnarray}
where the double integral can be simplified as
$\int^{\lambda_{\rm s}}_{\lambda_{\rm o}} {\rm d} \lambda' \int^{\lambda'}_{\lambda_{\rm o}} 
f(\lambda) {\rm d} \lambda =\int^{\lambda_{\rm s}}_{\lambda_{\rm o}} f(\lambda) (\lambda_{\rm s}-\lambda) 
{\rm d}\lambda$, and we use $e^{\theta i} g^{(1)}_{\sigma \rho,i}= 
g^{(1)}_{\sigma \rho ,\theta} /r$.

An analogous calculation gives
\begin{eqnarray}
\delta \varphi &=&-\frac{1}{r_{\rm o} \sin\theta_{\rm o}}\int^{\rm s}_{\rm o} {\rm d}\lambda 
\bigg(e^{\varphi i}[g^{(1)}_{0 i}k^0+g^{(1)}_{j i}k^j]^{\rm s}_{\rm o} \nonumber\\
&&- \frac{\lambda_{\rm s}-\lambda}{2r \sin \theta_{\rm o}} 
g^{(1)}_{\sigma\rho,\varphi}k^{\sigma} k^{\rho}  \bigg). 
\end{eqnarray}
At linear order in perturbation theory, we are allowed to identify $r$ and $r_{\rm o}$,  and 
$\theta$ with $\theta_{\rm o}$ inside the expressions, as those differences are of higher order.

\section{\label{App:Lum} Luminosity distance}

In this section we provide some essential steps for deriving the luminosity distance
at linear order. We follow closely Sasaki ~\cite{Sasaki1987}.
After gauge fixing our final expression agrees with Bonvin et al. ~\cite{Bonvin2006}.

The luminosity distance can be expressed as 
\begin{equation}
 D_L = \frac{\mathscr{A}_{\rm s} \omega_{\rm s}}{\mathscr{A}_{\rm o} \omega_{\rm o} } R_{\rm s}
\end{equation}
where $\mathscr{A}$ is the amplitude in the eikonal approximation of geometric optics.
According to the energy-momentum conservation 
and the geodesic equation, 
\begin{equation}
\nabla_\mu(\mathscr{A}^2 \hat{k}^\mu)=2\mathscr{A}(\frac{{\rm d}
\mathscr{A}}{{\rm d}\hat{\lambda}}+\frac{1}{2}\mathscr{A} \hat{\vartheta})=0,
\end{equation}
where $\hat{\vartheta}\equiv \nabla_\mu\hat{k}^\mu $.

In the conformally related geometry, one can verify that
\begin{equation} \label{eq:A}
 \nabla_\mu(\mathscr{A}^2 a^2 k^\mu)=2\mathscr{A}(\frac{{\rm d} (\mathscr{A}a)}{{\rm d} \lambda}
 +\frac{1}{2}\mathscr{A}a \vartheta)=0, 
\end{equation}
where $\vartheta\equiv \nabla_\mu k^\mu$ is the expansion of the congruence. 
The evolution of $\vartheta$ is described by its covariant derivative
along the null path, 
\begin{equation}
 \frac{{\rm d} \vartheta}{{\rm d} \lambda}=-R_{\mu\nu}k^\mu k^\nu
-\frac{1}{2}\vartheta^2-2\sigma^2.
\end{equation}
At the zeroth order, the Ricci tensor in the conformally related geometry $R_{\mu\nu}=0$, 
one simply gets
\begin{eqnarray}
 &&\frac{{\rm d} \bar{\vartheta}}{{\rm d}\lambda}+\frac{1}{2}\bar{\vartheta}^2=0 ,\nonumber\\
 && \bar{\vartheta}=\frac{2}{\lambda + c}.
\end{eqnarray}
We define $\lambda_{\rm o}$ and $\lambda_{\rm s}+\Delta\lambda_{\rm s}$ for the affine parameter
at observer and source, respectively. As shown in Fig.~\ref{fig:affine}, we
assume the source is spherical and its radius in terms of the affine parameter is
$\Delta\lambda_{\rm s}$.   
At the source $\bar{\vartheta} \to \infty$, then $c=-\lambda_{\rm s}-\Delta
\lambda_{\rm s}$, therefore to zeroth order
\begin{equation}
 \bar{\vartheta}=\frac{2}{\lambda  -\lambda_{\rm s} -\Delta \lambda_{\rm s}}.
\end{equation}
At first order,
\begin{equation}\label{eq:amp}
 \frac{{\rm d} \delta\vartheta}{{\rm d} \lambda}=-\delta R_{\mu\nu}k^\mu k^\nu
-\bar{\vartheta}\delta\vartheta. 
\end{equation}
Integration of Eq.~(\ref{eq:amp}) with the boundary condition
$\delta\vartheta(\lambda_{\rm s})=0$ yields
\begin{equation}\label{eq:expansion}
 \delta\vartheta(\lambda)=\frac{1}{(\lambda  -\lambda_{\rm s} - \Delta
\lambda_{\rm s})^2}\int^{\lambda_{\rm s}}_{\lambda} 
(\lambda'  -\lambda_{\rm s}-\Delta \lambda_{\rm s})^2 \delta R_{\mu\nu}k^\mu k^\nu {\rm d} \lambda'.
\end{equation}
According to Eq.~(\ref{eq:A}) 
\begin{equation}
 \mathscr{A}a=c_1 \exp[-\int_{\lambda_{\rm o}}^\lambda \frac{\vartheta}{2}{\rm d}\lambda].
\end{equation}
Therefore
\begin{equation}
 \frac{\mathscr{A}(\lambda_{\rm s}) a(\lambda_{\rm s})
}{\mathscr{A}(\lambda_{\rm o})a(\lambda_{\rm o})} 
=\frac{\lambda_{\rm s}-\lambda_{\rm o}+\Delta\lambda_{\rm s}}{\Delta\lambda_{\rm s}}\exp[-\int_{
\lambda_{\rm o}}^{\lambda_{\rm s}} \frac{\delta\vartheta}{2}{\rm d}\lambda].
\end{equation}
In the local inertial frame of the source $(\tilde{\eta},\tilde{x}^i)$, 
\begin{equation}
 \omega=g_{\mu\nu}u^\mu
\hat{k}^\nu=\frac{-1}{a^2}\frac{{\rm d}\tilde{
\eta}}{{\rm d}\lambda},
\end{equation}
and thus
\begin{equation}
 R_{\rm s} =\sqrt{\delta_{ij}{\rm d}\tilde{x}^i{\rm d}\tilde{x}^j}=|\Delta\tilde{\eta}|=a^2_{\rm s}
\Delta\lambda_{\rm s} \omega_{\rm s}.
\end{equation}
In the limit $\Delta\lambda_{\rm s}\to 0$, the  luminosity distance is
\begin{eqnarray}\label{eq:lum1}
D_{\rm L}&=&(1+z)\frac{\mathscr{A}_{\rm s} }{\mathscr{A}_{\rm o} }
R_{\rm s} \nonumber\\
 &=&a_{\rm o} (1+z) \frac{a_{\rm s} \omega_{\rm s}}{k^0}
(\eta_{\rm s}-\eta_{\rm o})[1-\frac{1}{\eta_{\rm s}-\eta_{\rm o}}\int^{\lambda_{\rm s}}_{\lambda_{\rm o}} l^0
{\rm d}\lambda \nonumber\\
 &&-\int_{\lambda_{\rm o}}^{\lambda_{\rm s}} \frac{\delta\vartheta}{2} {\rm d}\lambda] 
 \end{eqnarray}
where the term proportional to $l^0$ comes from replacing the affine parameter by the 
conformal time. At leading order we can further write 
\begin{equation}
\frac{1}{\eta_{\rm s}-\eta_{\rm o}}\int^{\lambda_{\rm s}}_{\lambda_{\rm o}} l^0
{\rm d} \lambda
=\frac{1}{\lambda_{\rm s} -\lambda_{\rm o}}\int^{\lambda_{\rm s}}_{\lambda_{\rm o}}\frac{l^0}{k^0}  {\rm d} \lambda.
\end{equation}
Using integration by parts,
\begin{eqnarray}
&&\int^{\lambda_{\rm s}}_{\lambda_{\rm o}} l^0   {\rm d}\lambda =\int^{\lambda_{\rm s}}_{\lambda_{\rm o}}
 (\lambda_{\rm s}-\lambda)\frac{ {\rm d}
l^0}{ {\rm d}\lambda} {\rm d}\lambda +(\lambda_{\rm s}-\lambda_{\rm o})l^0(\lambda_{\rm o}).\nonumber 
\end{eqnarray} 
According to Eq.~(\ref{equ:l0}) 
\begin{eqnarray}\label{equ:l}
&&\int^{\lambda_{\rm s}}_{\lambda_{\rm o}} {\rm d}\lambda (\lambda_{\rm s}-\lambda)\frac{{\rm d}
l^0}{ {\rm d}\lambda} \nonumber\\
&=&\int^{\lambda_{\rm s}}_{\lambda_{\rm o}} 
(\lambda_{\rm s}-\lambda)k^2\bigg[\dot{\Phi}-\dot{\Psi}-\frac{1}{2}e^ie^j(U_{i,j}+U_{j
,i}+\dot{h}_{ij})\bigg]{\rm d}\lambda\nonumber\\
&&-2k^0\int^{\lambda_{\rm s}}_{\lambda_{\rm o}} \bigg[\phi-(\ddot{E}+\dot{B})\bigg]
{\rm d}\lambda-[\dot{E}+B]|^{\rm s}_{\rm o} \nonumber\\
&&+(\lambda_{\rm s}-\lambda_{\rm o})[2k^0\phi_{\rm o}-k^0(\ddot{E}_{\rm o}+\dot{B}_{\rm o})+k^i(\dot{E}
_{\rm o}+B_{\rm o})_{,i}]\nonumber\\
\end{eqnarray} 
Inserting Eq.~(\ref{eq:expansion}) into the last term of Eq.~(\ref{eq:lum1}), and
integrating by parts, 
\begin{eqnarray}\label{eq:exp1}
&& \int_{\lambda_{\rm o}}^{\lambda_{\rm s}}
\frac{\delta\vartheta}{2} {\rm d}\lambda=\int^{\lambda_{\rm s}}_{\lambda_{\rm o}}
\frac{(\lambda_{\rm s}-\lambda)(\lambda-\lambda_{\rm o})}{2(\lambda_{\rm s}-\lambda_{\rm o})}\delta
R_{\mu\nu}k^\mu k^\nu  {\rm d}\lambda \nonumber\\
&=&\int^{\lambda_{\rm s}}_{\lambda_{\rm o}} {\rm d}\lambda
\frac{(\lambda_{\rm s}-\lambda)(\lambda-\lambda_{\rm o})}{2(\lambda_{\rm s}-\lambda_{\rm o})}k^2\bigg[
\Delta[\Phi-\Psi]-\Delta U_i e^i \nonumber\\
&&-[\Phi-\Psi]_{,ij}e^ie^j 
+\frac{1}{2}[\dot{U}_{i,j}+\dot{U}_{j,i}+\ddot{h}_{ij}-\Delta
h_{ij}]e^ie^j\bigg]\nonumber\\
&&-\psi_{\rm s}-\psi_{\rm o}+\frac{2}{\lambda_{\rm s}-\lambda_{\rm o}}\int^{\lambda_{\rm s}}_{\lambda_{\rm o}}\psi
 {\rm d}\lambda
\end{eqnarray}
Recall that the photon frequency $\omega_{\rm s}$ is
\begin{eqnarray}
 \omega_{\rm s}=g_{\mu\nu}u^\mu
\hat{k}^\nu=\frac{1}{a_{\rm s}}[-k^0-k^0\phi +k^i (v_i-B_{,i}-S_i)-l^0_{\rm s}]\nonumber\\
\end{eqnarray}
Finally, inserting Eq.~(\ref{eq:exp1}) and Eq.~(\ref{equ:l}) into Eq.~(\ref{eq:lum1}),
the luminosity distance can be expressed in terms of gauge
invariant quantities as
\begin{widetext}
\begin{eqnarray}\label{eq:lumm}
&&D_{\rm L}=a_{\rm o}(1+z)(\eta_{\rm o}-\eta_{\rm s}) \bigg[1+\psi_{\rm o}-\Phi_{\rm s}+\Psi_{\rm s} +e^i
(V_i)_{\rm s} +k^0\int^{\lambda_{\rm s}}_{\lambda_{\rm o}}(\dot{\Phi}-\dot{\Psi}) {\rm d}\lambda+\frac{2}{
\eta_{\rm s}-\eta_{\rm o}}\int^{\lambda_{\rm s}}_{\lambda_{\rm o}}k^0(\Phi-\Psi)  {\rm d}\lambda  \nonumber\\
&&+\frac{[\dot{E}+B]|^{\rm s}_{\rm o}}{\eta_{\rm s}-\eta_{\rm o}}
-\frac{1}{\eta_{\rm s}-\eta_{\rm o}}\int^{\lambda_{\rm s}}_{\lambda_{\rm o}} 
(\lambda_{\rm s}-\lambda)k^2\big(\dot{\Phi}-\dot{\Psi} \big) {\rm d}\lambda -\int^{\lambda_{\rm s}}_{\lambda_{\rm o}} {\rm d}\lambda
\frac{(\lambda_{\rm s}-\lambda)(\lambda-\lambda_{\rm o})}{2(\lambda_{\rm s}-\lambda_{\rm o})}k^2\bigg[
\Delta(\Phi-\Psi)-(\Phi-\Psi)_{,ij}e^ie^j\bigg]  \nonumber\\
&&-k^0\int^{\lambda_{\rm s}}_{\lambda_{\rm o}}\frac{1}{2}e^ie^j(U_{i,j}+U_{j,i}+\dot{h}_{ij}
) {\rm d}\lambda +\frac{1}{\eta_{\rm s}-\eta_{\rm o}}\int^{\lambda_{\rm s}}_{\lambda_{\rm o}} 
(\lambda_{\rm s}-\lambda)k^2\big[\frac{1}{2}e^ie^j(U_{i,j}+U_{j,i}+\dot{h}_{ij})\big]
{\rm d}\lambda  \nonumber\\
&&-\int^{\lambda_{\rm s}}_{\lambda_{\rm o}} {\rm d}\lambda
\frac{(\lambda_{\rm s}-\lambda)(\lambda-\lambda_{\rm o})}{2(\lambda_{\rm s}-\lambda_{\rm o})}k^2\big[
\frac{1}{2}(\dot{U}_{i,j}+\dot{U}_{j,i}+\ddot{h}_{ij}-\Delta
h_{ij})e^ie^j-\Delta U_ie^i\big]\bigg].
\end{eqnarray}
\end{widetext}

\begin{acknowledgments}
We thank Camille Bonvin, Ruth Durrer, Matthias Rubart, Daniel Boriero and Jaiyul Yoo for valuable 
discussions. We acknowledge the support from the Research Training Group 1620 ``Models of Gravity'' funded by Deutsche Forschungsgemeinschaft. 
\end{acknowledgments}

\bibliographystyle{apsrev4-1}
\bibliography{diffNCv5.bib}

\end{document}